\documentclass[pdflatex,sn-mathphys-num]{sn-jnl}

\usepackage{graphicx}%
\usepackage{multirow}%
\usepackage{amsmath,amssymb,amsfonts}%
\usepackage{amsthm}%
\usepackage{mathrsfs}%
\usepackage[title]{appendix}%
\usepackage{xcolor}%
\usepackage{textcomp}%
\usepackage{manyfoot}%
\usepackage{booktabs}%
\usepackage{algorithm}%
\usepackage{algorithmicx}%
\usepackage{algpseudocode}%
\usepackage{listings}%

\usepackage{hyperref} 
\providecommand{\burl}[1]{\url{#1}}

\theoremstyle{thmstyleone}%
\theoremstyle{thmstyletwo}%

\theoremstyle{thmstylethree}%

\begin{document}

\title[From three-body resonances to bound states in a continuum: pole trajectories]{From three-body resonances to bound states in a continuum: pole trajectories}


\author*[1]{\fnm{Lucas} \sur{Happ}}\email{lucas.happ@riken.jp}

\affil*[1]{\orgdiv{Few-body Systems in Physics Laboratory}, \orgname{RIKEN Nishina Center for Accelerator-Based Science}, \city{Wak\={o}}, \state{Saitama} \postcode{351-0198}, \country{Japan}}


\abstract{We investigate the formation of three-body bound states in the continuum from resonances by tracing their pole trajectories in the complex energy plane under variation of system parameters. Using a one-dimensional model of two identical bosons and a distinguishable particle interacting via Gaussian potentials, we systematically vary the interaction strength, interaction range, and mass ratio. Our results confirm the single-resonance parametric nature of such three-body bound states in a continuum and extend this characterization to a broader parameter space. While all parameter variations lead to formation of such states, the mass ratio exhibits a richer structure with multiple locations arranged in regular oscillatory trajectories. All three parameters show width minima at a common value of the relative momentum between outgoing subsystems, suggesting some robustness in the formation mechanism.}




\maketitle

\section{Introduction}

One of the core motivations of few-body physics is to help understand many-body phenomena beyond single-particle descriptions. Examples include three-body recombination losses in Bose-Einstein condensates~\cite{esry1999,kraemer2006} and three-body forces~\cite{endo2025} in nuclear structure. However, intrinsic interest in few-body systems as natural extensions of exactly solvable models arose already in the early days of quantum mechanics, as demonstrated by studies on the H$_2^+$ molecule~\cite{burrau1927} and the triton~\cite{bethe1936}. The Thomas collapse~\cite{thomas1935} in the triton and the more famous Efimov effect~\cite{efimov1970,naidon2017} in general three-body systems, together with the formal description of three-body scattering by Faddeev~\cite{faddeev1961}, helped establish the field of few-body physics. Since it concerns itself with the number of particles rather than specific constituents or interactions, few-body physics is intrinsically interdisciplinary and naturally spans from high-energy regimes such as hadron~\cite{guo2018,arifi2024} and nuclear physics~\cite{nielsen2001}, to low-energy physics as in semiconductors~\cite{kazimierczuk2014,belov2024} and ultracold atoms~\cite{kraemer2006,pires2014}.

While early work in few-body physics was performed primarily on bound states, the majority of few-body states in nature are metastable states, so-called resonances~\cite{kukulin1989,moiseyev1998}. This is because few-body states can often break up into their constituent parts, owing to the existence of one or more deeply-bound states of their subsystems. Most studies have focused on bound state properties or resonance positions, while the lifetimes and stability of few-body systems, despite their importance, remain relatively underexplored~\cite{penkov1999,nielsen2002,pricoupenko2010a,happ2024,happ2025}. This is of particular relevance, since it has been a long-standing challenge to stabilize three-body states such as Efimov trimers~\cite{laird2018}.

The field of ultracold atoms offers a promising platform to address such challenges due to the high tunability and controllability of these systems. One aspect demonstrating these features is the ability to use external trapping potentials to confine atoms to geometries of reduced dimensionality~\cite{bloch2005}, such as pancake-shaped (quasi-2D) or cigar-shaped (quasi-1D) configurations. Such experimental realizations~\cite{serwane2011,zurn2012,murmann2015} have motivated several theoretical studies on low-dimensional few-body systems~\cite{dodd1970,mehta2005,mehta2007,kartavtsev2009,blume2012,pricoupenko2018,sowinski2019,happ2019,happ2021,happ2022,mistakidis2023,schnurrenberger2025}.

In a previous work~\cite{happ2024}, we studied three-body resonances in a mass-imbalanced one-dimensional three-body system and found a strong dependence of their lifetimes on the mass ratio between the species. More importantly, in a subsequent study employing a two-channel model, we identified the underlying mechanism through which the lifetime of general three-body resonances can be increased up to theoretically infinity~\cite{happ2025}. In this case, the resonances do not decay and turn into stable states, so-called bound states in the continuum~\cite{hsu2016}. These peculiar states have been studied since the early days of quantum mechanics, famously by von Neumann and Wigner~\cite{vonNeumann1929}, and to the present day they have been realized experimentally in several systems, predominantly in photonics~\cite{marinica2008,pankin2020}.

In the present article, we study three-body resonances in the same one-dimensional system as in Ref.~\cite{happ2024}, but this time under the aspect of pole trajectories and under variation of several system parameters including the strength and range of the interaction. Pole trajectories, which are commonly employed in nuclear and particle physics~\cite{collins1968,glockle1978,badalyan1982,matsuyama1991,badalian2002}, are a tool to visualize how the real and imaginary parts of a resonance energy are related as system parameters vary. This approach allows us to systematically compare the influence of different parameters on the resonance properties and to trace the continuous transformation from resonances to bound states in the continuum. We investigate the role of the relative momentum between subsystem fragments, and trace pole trajectories in the complex energy plane. Numerical convergence and the complex-scaled spectrum are discussed in the appendices.

\section{Model and Method}

\subsection{Three-body Hamiltonian and interactions} \label{sec:hamiltonian}

In this work we consider a three-body system of two identical bosons of masses $m_B$, and a third, distinguishable particle of mass $m_X$, all confined to one spatial dimension. The parameter $\beta = m_B/m_X$ denotes the mass ratio. We assume the interaction between different particles to be of Gaussian shape
\begin{equation}\label{eq:pot}
    V(r) = v_0 \exp\left[-\mu_g (r/r_0)^2\right], \qquad v_0 <0
\end{equation}
and no interaction between the two identical bosons. This reduces the number of system parameters to a minimum and allows to study the remaining ones ($v_0,\mu_g,\beta$) in a more isolated manner. 

We work in dimensionless units throughout this paper, where masses are given in units of $m_X$, distances in units of a characteristic length scale $r_0$, and energies in units of $\hbar^2/(m_X r_0^2)$. Transforming to the center-of-mass frame of the three-body system and separating the relative dynamics from the center-of-mass motion, the three-body system is governed by the Schrödinger equation
\begin{equation}\label{eq:3bsgl}
    \left[-\frac{1}{2\mu_{12}}\frac{\partial^2}{\partial r_3^2} -\frac{1}{2\mu_{12,3}}\frac{\partial^2}{\partial R_3^2} + V\left(r_2\right) + V\left(r_3\right) - E^{(3)}\right] \psi(r_3,R_3) = 0.
\end{equation}
We use the index $1$ for the different particle, and $2,3$ for the two identical Bosons. Then, $r_3$ ($r_2$) denotes the distance between the pair of particles $\{1,2\}$ ($\{3,1\}$), and $R_3$ the distance between particle $3$ relative to the center-of-mass of the two-particle subsystem of particles $\{1,2\}$. The corresponding reduced masses are defined as
\begin{equation}
    \mu_{12} = \frac{m_1 m_2}{m_1 + m_2} = \frac{\beta}{\beta+1},\qquad \mu_{12,3} = \frac{m_3 (m_1 + m_2)}{m_1 + m_2 + m_3} = \frac{\beta(1+\beta)}{1+2\beta}.
\end{equation}

Since in this work we are interested in three-body resonances, we employ the complex scaling method (CSM)~\cite{moiseyev1998}. This method transforms resonances, which appear as poles in the complex energy plane~\cite{schleich2023}, into discrete eigenvalues that can be computed using standard bound-state techniques. We then solve the three-body Schrödinger equation, Eq.~\eqref{eq:3bsgl}, with complex scaling via the Gaussian expansion method (GEM)~\cite{hiyama2003}. This method relies on expanding the total three-body state into a superposition of centered Gaussian functions with different widths for each of the Jacobi sets $k = 1,2,3$ (for our system we only need $k=2$ and $k=3$) and for each of the relative coordinates $r_k$, $R_k$ in each set. This coherent superposition of Gaussians allows to represent also the more general class of non-Gaussian states~\cite{happ2018,walschaers2021}. We use the open source julia package FewBodyToolkit.jl~\cite{happ2025a} which provides a general implementation of the GEM for few-body systems, with options to enable the CSM. With this, we obtain the complex scaled spectrum from which we can extract the complex energies of the three-body resonances. An example spectrum is shown in Appendix~\ref{app:spectrum}.

\subsection{Threshold structure}

\begin{figure}
    \centering
    \includegraphics[width=0.6\linewidth]{ 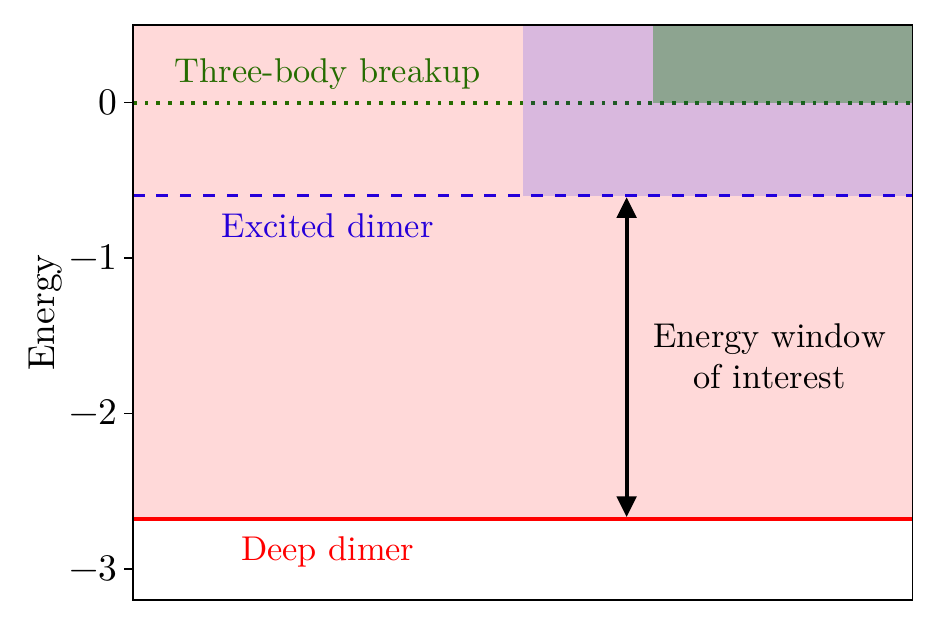}
    \caption{Energy diagram showing the relevant thresholds: deep dimer $E_0^{(2)}$ (red), excited dimer $E_1^{(2)}$ (blue), three-body breakup $E^{(3)}=0$ (green), and associated continua (shaded areas). Within this work we focus on three-body resonances located between the two dimer states, which have only one open decay channel: to decay into the 2+1 continuum of the deep dimer and a free boson.}
    \label{fig:thresholds}
\end{figure}

We define the zero of energy at the three-body breakup threshold $E^{(3)} = 0$. For the parameter values considered in this work, the two-body subsystem consisting of the distinguishable particle and one boson supports two bound states: a deeply-bound ``deep dimer'' state with binding energy $E_0^{(2)}$, and an ``excited dimer'' state with binding energy $E_1^{(2)}$. The three-body resonances of interest lie in the energy window $E_0^{(2)} < \mathrm{Re}\,E^{(3)} < E_1^{(2)}$ between the two dimer states and can decay into the $2+1$ continuum of the deep dimer plus a free boson. This constitutes therefore their only open decay channel. Figure~\ref{fig:thresholds} illustrates the threshold structure. Three relevant thresholds at (i) the deep dimer energy $E_0^{(2)}$, (ii) the excited dimer energy $E_1^{(2)}$, and (iii) the three-body breakup $E^{(3)} = 0$ each induce a corresponding continuum (red, blue, and green shaded areas).

Resonances appear as poles on the (unphysical) second Riemann sheet with complex energies 
\begin{equation}
    E^{(3)} = \mathrm{Re}\,E^{(3)} + i\mathrm{Im}\,E^{(3)} \equiv E_R + i\,E_I, \qquad \Gamma = -2E_I
\end{equation}
with $\Gamma > 0$. Bound states in the continuum~\cite{hsu2016} are special cases of resonances with zero width ($\Gamma = 0$) and thus lie on the real axis. Since they are normalizable eigenstates with real eigenvalues above the threshold, they represent the limiting case where a second-sheet resonance pole reaches the real axis. From $E_0^{(2)}$ to $+\infty$ on the real axis lies the branch cut  which corresponds to the 2+1 continuum of the deeply bound dimer state and a free particle. There are two more branch cuts along the real axis, one starting from $E_1^{(2)}$ (2+1 continuum of excited dimer plus free boson), and another one starting from $E^{(3)} = 0$ (continuum of three free particles). An example spectrum within the CSM, featuring bound states, resonances and rotated continua (branch cuts) is shown in Appendix \ref{app:spectrum}.

\section{Pole trajectories and BIC formation}
\subsection{Parameter dependence of energies and widths}

\begin{figure}[htbp]
    \centering
    \includegraphics[width=\linewidth]{ 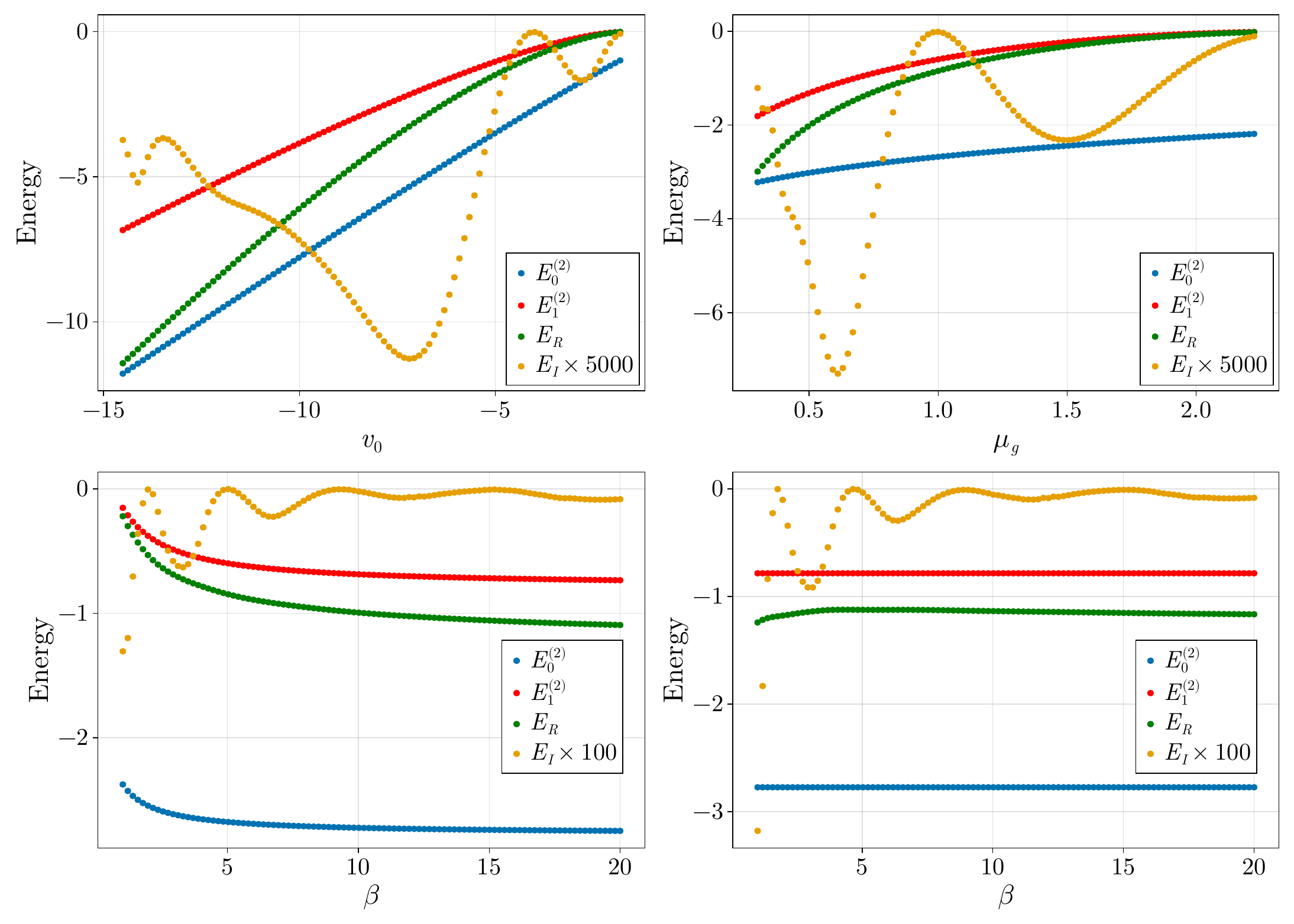}
    \caption{Energies as a function of system parameters $v_0$ (top left panel), $\mu_g$ (top right panel), and $\beta$ (bottom left panel). The bottom right panel shows the same $\beta$ dependence, but the energies are scaled in units of $\hbar^2/2\mu_{12}r_0^2$ instead, such that the dimer energies are constant. In particular, we present the deep dimer energy $E_0^{(2)}$ (blue), the excited dimer energy $E_1^{(2)}$ (red), the three-body resonance position $E_R$ (green) and its width via $E_I$ (yellow).
    }
    \label{fig:energies}
\end{figure}

We limit our analysis to the parameter regime where the three-body resonance lies between the two dimer states, as indicated in Fig.~\ref{fig:thresholds}. Figure~\ref{fig:energies} displays both two-body binding energies, $E_0^{(2)}$ and $E_1^{(2)}$, the three-body resonance position $E_R$, and the imaginary part $E_I$ as functions of the three system parameters $v_0$, $\mu_g$, and $\beta$.

The real parts of all energies exhibit monotonic behavior as expected from physical intuition\footnote{The exception is the bottom-right panel where due to a different scaling, $E_R$ first increases for small mass ratios, before it eventually slowly decreases.}: more attractive interactions lead to more negative binding energies. Specifically, for increasing $|v_0|$ (stronger attraction) or decreasing $\mu_g$ (wider potential), both dimer and trimer states become more bound. Similarly, increasing the mass ratio $\beta$ leads to deeper binding. In contrast, for all three parameters the resonance width (expressed via $E_I = -\Gamma/2$) shows highly nonlinear behavior.

Most notably, for each of the three parameter variations, we observe at least one point where the width vanishes, indicating the formation of a bound state in the continuum. For the interaction parameters $v_0$ and $\mu_g$, we find a single such point within the accessible parameter range, while the mass ratio variation exhibits a richer structure with multiple zeros of the width. The mechanism underlying this stabilization was analyzed in Ref.~\cite{happ2025} based on a two-channel description, and was identified as a tunable vanishing of a transition element. Since there is only a single continuum and a single resonance involved, these states are classified as single-resonance parametric BICs. Figure~\ref{fig:width_log} in Appendix~\ref{app:width} shows the width on a logarithmic scale, demonstrating that at these parameter values it indeed approaches zero, up to numerical precision.

The energy difference $E^{(3)} - E_0^{(2)}$ between the resonance and the lower threshold might intuitively be expected to give a simple estimate for the resonance width. However, the non-monotonic structure of $\Gamma$ observed here suggests a more complex dependence. This is further highlighted in the bottom right panel of Fig.~\ref{fig:energies}, where we show results for a mass ratio scan in rescaled units such that the dimer energies stay constant~\cite{happ2024}. Here, despite the nearly constant resonance position $E_R$ and henceforth nearly constant energy difference, the width still exhibits strong variations.

\subsection{Analysis of the relative momentum}

In our previous work~\cite{happ2025}, we suggested that the relative momentum
\begin{equation} \label{eq:prel}
 p_\mathrm{rel} \equiv \sqrt{2\mu_{12,3}\left[E_R - E^{(2)}_0\right]}
\end{equation}
between the deep dimer and the unbound boson serves as a key control parameter for the stabilization mechanism. Here we can use it to simultaneously analyze the width-behavior across different parameter sweeps along a common axis.

\begin{figure}[htbp]
    \centering
    \includegraphics[width=0.49\linewidth]{ 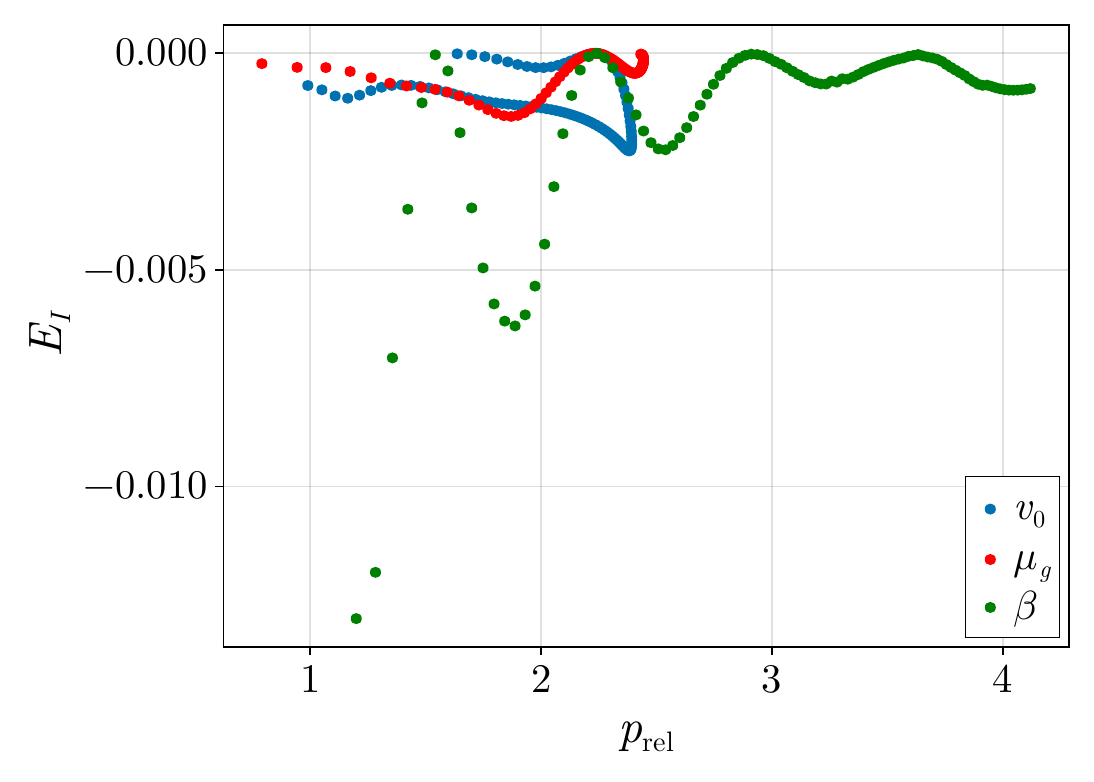}
    \includegraphics[width=0.49\linewidth]{ 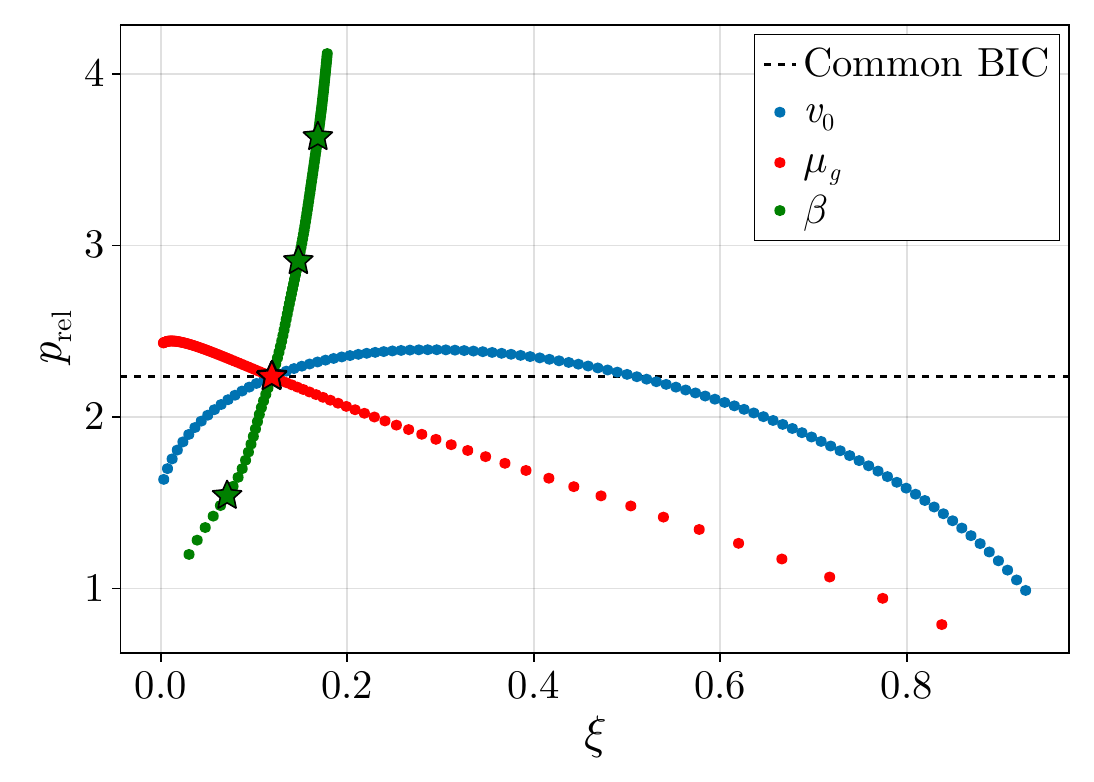}
    \caption{Importance of the relative momentum. Left panel: Imaginary part $E_I$ of the three-body resonance energy as a function of $p_{\text{rel}}$ for the three system parameters $v_0$ (blue), $\mu_g$ (red), and $\beta$ (green), showing minima near the common BIC location. Right panel: $p_{\text{rel}}$ as a function of the normalized parameter $\xi$, Eq.~\eqref{eq:xi}, under variation of the three system parameters. Values of $p_{\text{rel}}$ for which the width vanishes are marked by stars. The common BIC at $p_{\text{rel}} \approx 2.2$ is indicated by a horizontal dashed line. 
    }
    \label{fig:prel}
\end{figure}

In the left panel of Fig.~\ref{fig:prel} we display the imaginary part of the resonance energy as a function of the relative momentum. This reveals a notable feature: near $p_{\text{rel}} \approx 2.2$, all three parameter variations show minima in the width, suggesting some sort of universality and robustness in the BIC formation. The similar behavior around the common BIC location also indicates that further analytical insight can be gained. However, away from this common BIC location, the curves for $E_I$ diverge from each other, hence $p_{\text{rel}}$ does not seem to provide a global explanation. We also note a distinct u-turn in $p_\mathrm{rel}$ for the $v_0$ variation.

Since we trace the resonance pole as long as it remains between the two dimer energies, we introduce the dimensionless parameter
\begin{equation}\label{eq:xi}
    \xi \equiv \frac{E^{(2)}_1 - E_R}{E^{(2)}_1 - E^{(2)}_0}
\end{equation}
which by construction remains between $0$ and $1$ throughout our analysis. In the right panel we display the relative momentum as a function of $\xi$. We see that $p_{\text{rel}}$ itself exhibits non-monotonic dependence for $v_0$ and, to a lesser extent, for $\mu_g$ which is reflected in the u-turn shape in the left panel. This non-monotonicity can be understood from the behavior visible in Fig.~\ref{fig:energies}: the energy difference $E_R - E_0^{(2)}$ is non-monotonic because the trimer energy initially increases approximately quadratically near threshold before transitioning to a more linear regime, while the deep dimer energy already follows a linear trend throughout the parameter range. For $\mu_g$, this effect is less pronounced, and for the mass ratio $\beta$, the energy difference is approximately monotonic. We note that the mass ratio strongly affects the accessible range of $p_{\mathrm{rel}}$ through the reduced mass $\mu_{12,3}$, allowing it to reach much larger values compared to the interaction parameter scans.

\subsection{Pole Trajectories}
Via so-called pole trajectories we can trace a resonance location in the complex energy plane for various values of the system parameters. Here, we investigate trajectories under variation of the interaction strength $v_0$, the interaction range parameter $\mu_g$, and the mass ratio $\beta$ between the bosons and the distinguishable particle. The trajectories are plotted in the complex energy plane spanned by $E_R$ and $E_I$, hence the origin corresponds to $E^{(3)} = 0$. We note that throughout these parameter scans, the resonance pole does not cross any threshold or branch cut. The trajectory remains within the energy regime defined by the two dimer states (see Figs.~\ref{fig:thresholds} and~\ref{fig:energies}).

\begin{figure}[htbp]
    \centering
    \includegraphics[width=0.49\linewidth]{ 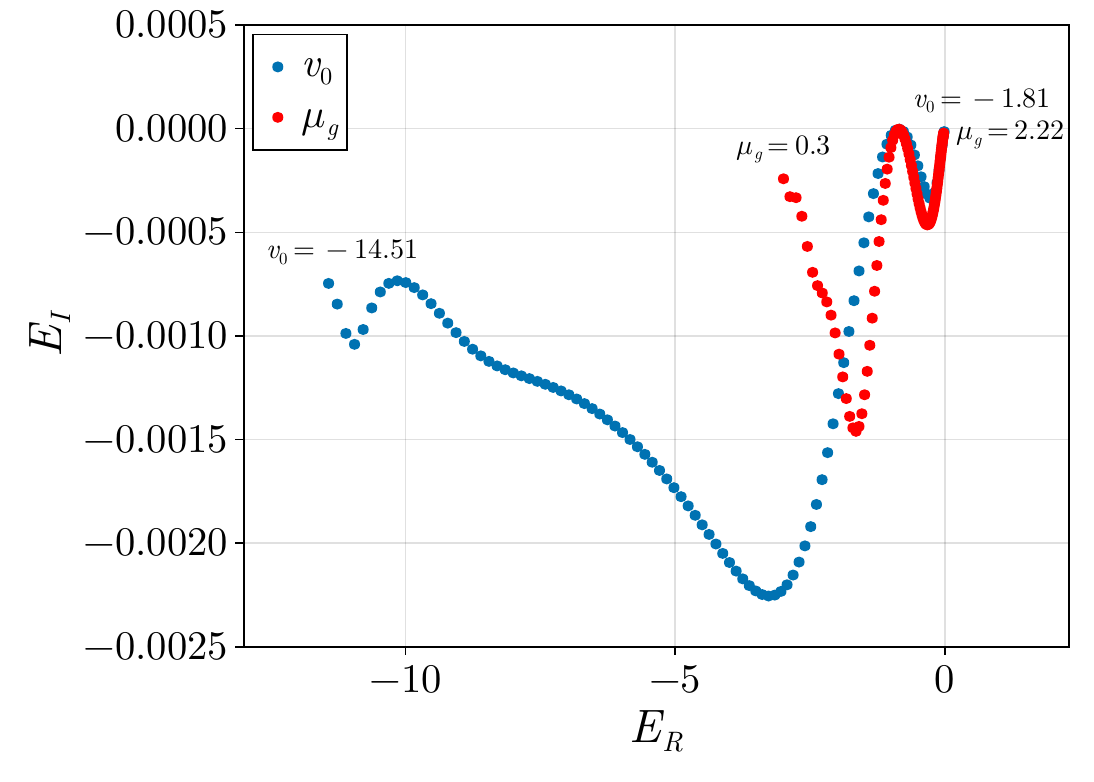}
    \includegraphics[width=0.49\linewidth]{ 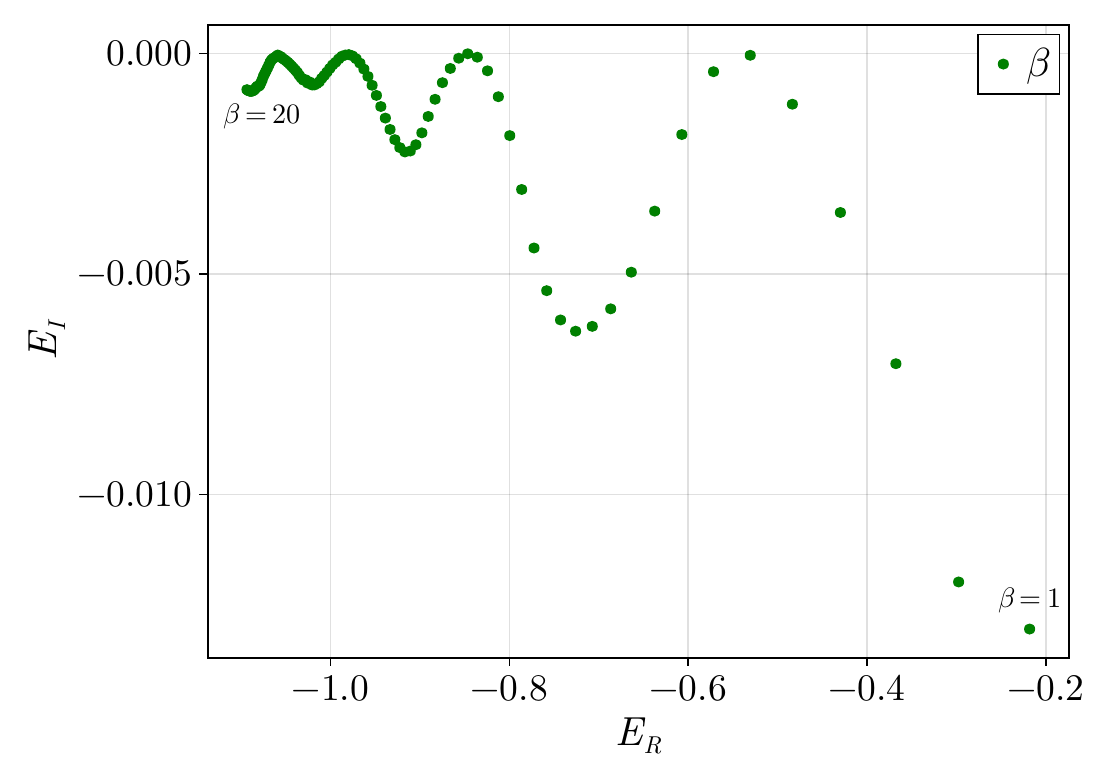}
    \caption{Trajectories of the three-body resonance pole in the complex energy plane for variations of the system parameters $v_0$ (blue) and $\mu_g$ (red), both in the left panel, and for $\beta$ (green) in the right panel. Left panel: For the $v_0$ variation, the parameter ranges from $v_0 = -1.81$ (right) to $v_0 = -14.51$ (left), and we keep $\mu_g = 1$ and $\beta = 5$ fixed. For the $\mu_g$ variation, the parameter ranges from $\mu_g = 2.22$ (right) to $\mu_g = 0.3$ (left), while $v_0 = -4$ and $\beta = 5$ are fixed. Right panel: $\beta$ is varied from $\beta = 1$ (right) to $\beta = 20$ (left), and $v_0 = -4$ and $\mu_g = 1$ are kept constant. Points where the trajectories touch the real axis correspond to BIC formations.}
    \label{fig:trajectories}
\end{figure}

Figure~\ref{fig:trajectories} shows the resulting pole trajectories whose behavior is consistent with the analysis in the previous subsections. The pole trajectory representation provides a direct visualization of the relationship between resonance position $E_R$ and width $\Gamma = -2E_I$. An important distinction from Fig.~\ref{fig:prel} is that in Fig.~\ref{fig:trajectories} the horizontal axis corresponds to $E_R$, which depends (mostly) monotonically on the parameters $v_0$, $\mu_g$, and $\beta$, whereas $p_{\text{rel}}$ showed non-monotonic behavior for the interaction parameters. Consequently, the pole trajectories presented here do not exhibit the intricate structure visible in Fig.~\ref{fig:prel} (right panel).

For the interaction parameters $v_0$ and $\mu_g$ (left panel), both variations produce qualitatively similar trajectories. The pole begins near the three-body breakup threshold ($E^{(3)} = 0$) and moves to lower energies as the interaction becomes more attractive. The width initially increases, then decreases to zero at the BIC, before increasing again. For both interaction parameters, the BIC occurs at nearly the same real energy, suggesting that this feature is relatively robust with respect to the specific way the interaction is modified. The overall trajectory shape is somewhat irregular, exhibiting a single point where the pole touches the real axis within the accessible parameter range.

The mass ratio variation (right panel) is displayed separately because the resonance position varies over a comparatively small energy interval. Here, the trajectory extends further away from the real axis, with imaginary parts nearly two orders of magnitude larger than for the interaction parameter scans. More strikingly, this trajectory displays a much more regular, oscillatory pattern, with the pole touching the real axis at multiple points, indicating the formation of several BICs as the mass ratio is varied.

\section{Conclusion and Outlook}

In summary, we have traced the formation of three-body bound states in the continuum from resonances by following pole trajectories in the complex energy plane as system parameters are varied. Using a one-dimensional model of two identical bosons and a distinguishable particle, we confirmed the single-resonance parametric nature of such BICs initially reported in our previous work~\cite{happ2025}, and extended this characterization to a broader selection of system parameters.

Our investigation reveals both similarities and differences in how interaction parameters (strength and range) and the mass ratio affect resonance properties. All three parameter variations lead to the formation of at least one BIC, however, the detailed behavior differs: variations of $v_0$ and $\mu_g$ produce single BICs within the accessible parameter ranges with somewhat irregular pole trajectories, while the variation of $\beta$ yields multiple BIC locations in a more regular, oscillatory pattern extending further into the complex plane.

A notable finding is that all parameter variations show a common BIC at the same relative momentum value, suggesting a degree of universality and robustness in the formation mechanism. Developing an analytical understanding of the behavior near that value as well as understanding to which extend our results are specific to the Gaussian potential remains an open question for future investigations.

\bmhead{Acknowledgments} We thank P.~Naidon, R.~Lazauskas, and T.~Kinugawa for fruitful discussions. L.~H. is supported by the RIKEN special postdoctoral researcher program.

\begin{appendices}
\section{Appendix}\label{app:width}

\subsection{Width in logarithmic scale}

\begin{figure}[htbp]
    \centering
    \includegraphics[width=0.6\linewidth]{ 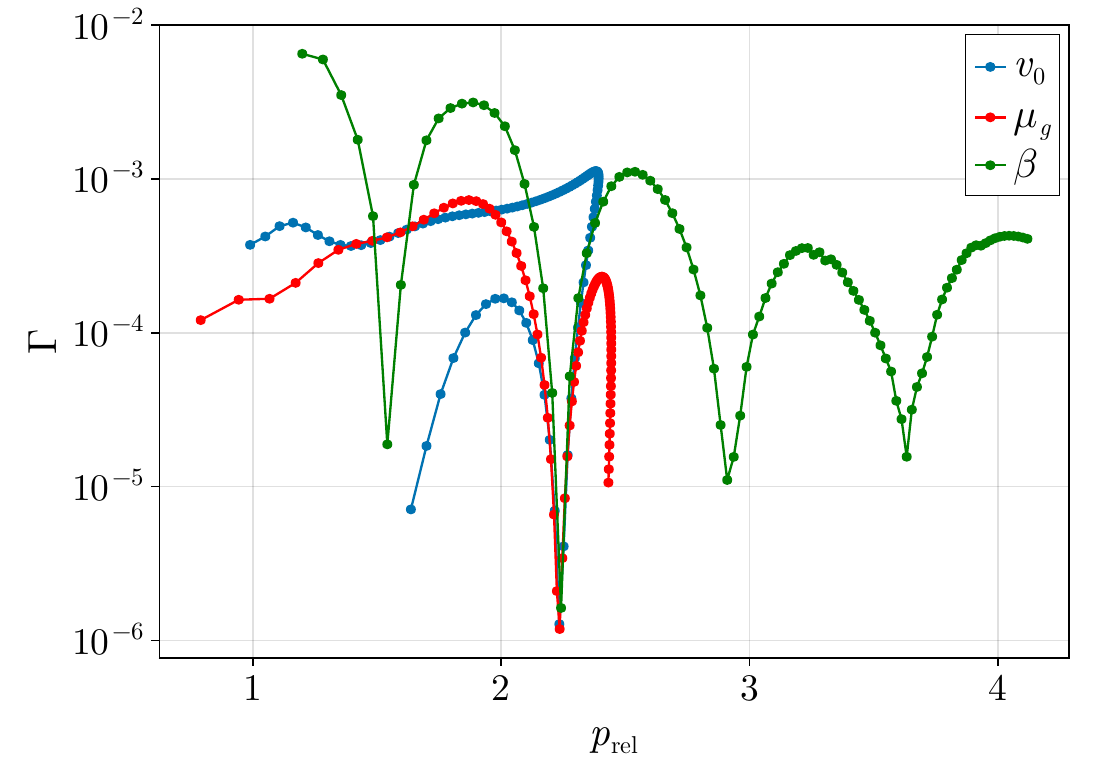}
    \caption{Width $\Gamma$ in log-scale as a function of the relative momentum $p_{\text{rel}}$, Eq.~\eqref{eq:prel}, for all three parameter variations: $v_0$ (blue), $\mu_g$ (red), and $\beta$ (green). The lower limit of the curves indicates the estimated numerical accuracy. All variations show at least one zero of the width, with a common BIC near $p_{\text{rel}} \simeq 2.2$.}
    \label{fig:width_log}
\end{figure}

Here, we examine the decay width $\Gamma = -2\text{Im}(E^{(3)})$ as a function of the system parameters on a logarithmic scale. Figure~\ref{fig:width_log} shows the width as a function of the relative momentum $p_{\text{rel}}$, Eq.~(5), for all three parameter variations: $v_0$ (blue), $\mu_g$ (red), and $\beta$ (green).

All curves reveal at least one zero in the decay width, confirming the existence of a three-body BIC for each parameter variation. One common BIC location occurs near $p_{\text{rel}} \simeq 2.2$, where the width drops close to $10^{-6}$ in the natural units of the problem. Away from this minimum, the width increases sharply by more than two orders of magnitude, demonstrating the dramatic enhancement in resonance lifetime at the BIC. The minimum widths obtained in these scans are limited by the numerical precision of our calculation and the coarseness of the parameter sampling. Finer parameter scans would likely reveal even deeper minima, approaching the numerical accuracy of the calculation, estimated to be around $10^{-6}$, see Appendix~\ref{app:conv}

\subsection{Numerical convergence}\label{app:conv}

\begin{figure}[htbp]
    \centering
    \includegraphics[width=0.6\linewidth]{ 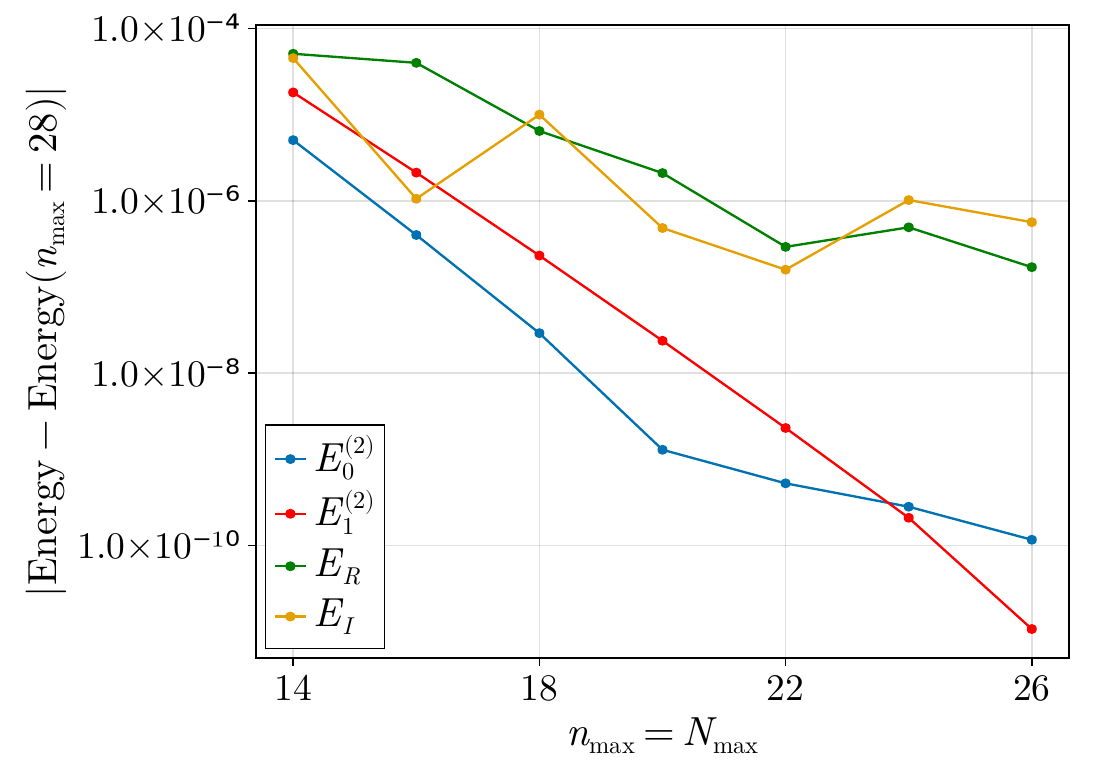}
    \caption{Convergence of energies with increasing Gaussian basis size $n_{\max} = N_{\max}$ for $v_0 = -4$, $\mu_g = 1$, and $\beta = 5$. In particular, we present the deep dimer energy $E^{(2)}_0$ (blue), the excited dimer energy $E^{(2)}_1$ (red), the three-body resonance position $E_R$ (green) and its width via $E_I$ (yellow). To display them in logarithmic scale we  subtracted them from their values at $n_{\max} = 28$ and took the absolute value. All quantities show good convergence for $n_{\max} \geq 26$ which we use for the calculations in this work.}
    \label{fig:convergence}
\end{figure}

To validate the numerical accuracy of our calculations, we perform convergence tests by varying the Gaussian basis size. Figure~\ref{fig:convergence} shows the convergence of both dimer binding energies and real- and imaginary parts of the three-body resonance energy as a function of the number $n_{\max} = N_{\max}$ of Gaussian basis functions in each Jacobi coordinate. The parameters used for this analysis are $v_0 = -4$, $\mu_g = 1$, and $\beta = 5$, corresponding to a configuration near a BIC where the width is small and numerical convergence is most demanding.

Both dimer energies show rapid and stable convergence with increased basis size whereas the three-body energies, as expected, converge more slowly. Nevertheless, we find good convergence starting from $n_{\max} = N_{\max} \gtrsim 26$. We therefore use these values for all calculations presented in this work. They provide converged results within the numerical precision of about $10^{-6}$ (compare lower limit of Fig. \ref{fig:width_log}).

The minimum and maximum range parameters controlling the widths of the Gaussian basis functions are chosen as $r_{\min} = 0.055, R_{\min} = 0.15$ and $r_{\max} = 15.2, R_{\max} = 16.59$ in dimensionless units, see section~\ref{sec:hamiltonian}. The ranges are distributed geometrically~\cite{hiyama2003,happ2025a} between these limits, providing adequate coverage of both short-range and long-range parts of the three-body wavefunction. Results were found to be insensitive to small variations in these parameters.

\subsection{Complex-scaled spectrum}\label{app:spectrum}

\begin{figure}[htbp]
    \centering
    \includegraphics[width=0.49\linewidth]{ 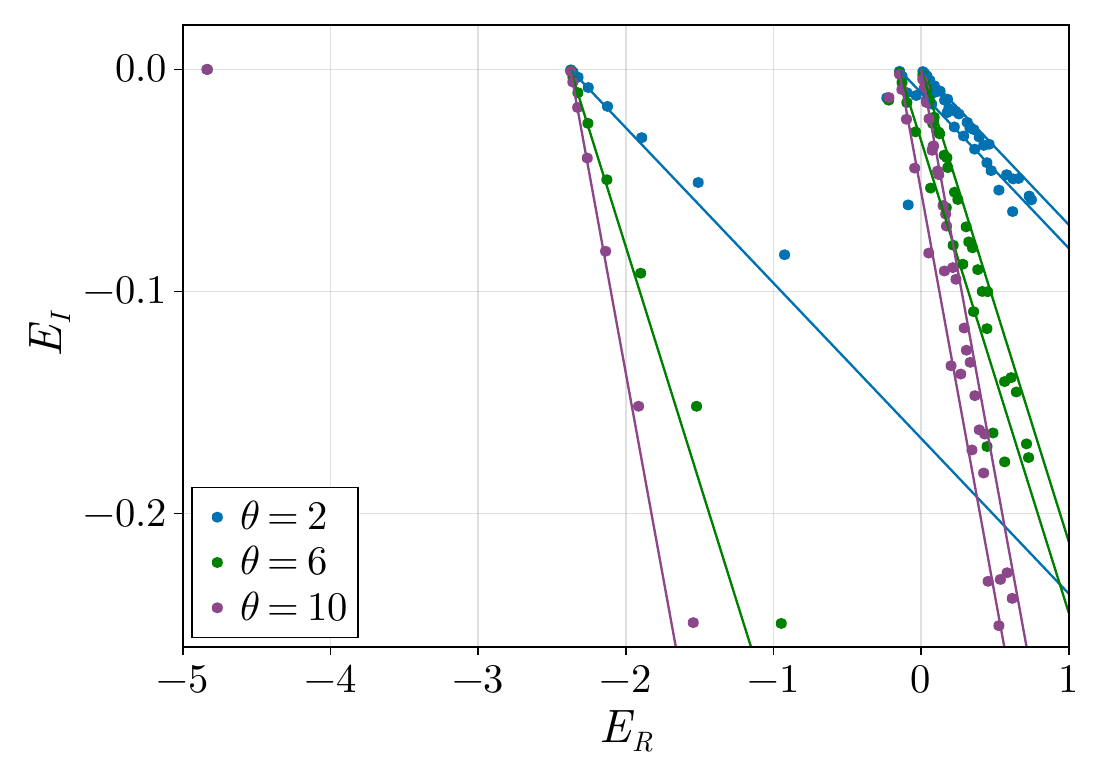}
    \includegraphics[width=0.49\linewidth]{ 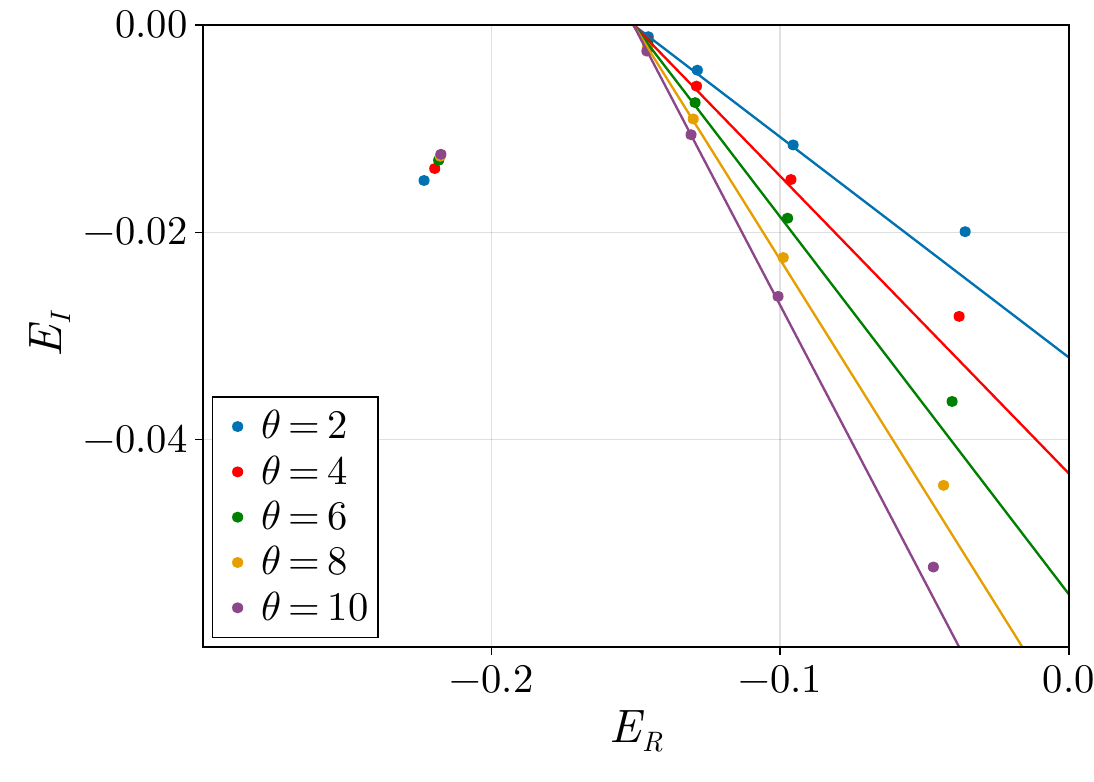}
    \caption{Complex-scaled energy spectrum for $v_0 = -4$, $\mu_g = 1$, and $\beta = 1$ at different rotation angles $\theta$ (see legend). The left panel shows the full spectrum with a bound state near $E \approx -5$, the rotated continuum, and a resonance near $E \approx -0.2 - 0.1i$. The right panel provides a magnified view of the resonance region, demonstrating the convergence with $\theta$ while the continuum rotates with changing angle.}
    \label{fig:spectrum}
\end{figure}

Figure~\ref{fig:spectrum} displays the complex-scaled energy spectrum obtained from the complex scaling method for the parameter values $v_0 = -4$, $\mu_g = 1$, and $\beta = 1$. The left panel shows the full spectrum, revealing several characteristic features: a true three-body bound state appears near $E \approx -5$ on the real axis, the continuum states are rotated downward into the complex plane, forming dense bands along rays at angles $-2\theta$ from the real axis~\cite{moiseyev1998}, and we can also identify a three-body resonance near $E \approx -0.2 - 0.1i$.

The right panel shows a magnified view focusing on the resonance region, demonstrating the stability of the resonance eigenvalue with respect to variations in the rotation angle: while the continuum rotates downward with changing $\theta$, the resonance position shows only little variation. This $\theta$-independence is the defining characteristic of resonance poles~\cite{moiseyev1998} and validates our numerical approach. For all calculations in this work, we employ a rotation angle of $\theta = 8^\circ$.

Within the complex scaling formalism, resonances become accessible as discrete eigenvalues because the method rotates the branch cut downward in the complex plane, effectively exposing poles that would otherwise lie on the unphysical second Riemann sheet. We emphasize that complex scaling is a numerical technique to make resonances tractable using standard bound-state methods. The physical interpretation of resonances as second-sheet poles remains unchanged.

\end{appendices}

\bibliography{PoleTrajectories}

\end{document}